\documentclass[10pt,letterpaper,twocolumn]{article} 
\usepackage{ol2}
\usepackage{psfrag}
\usepackage{textcomp}
\usepackage{amsmath}
\usepackage{ifthen}
\begin{document}
\twocolumn[ 
%
%
\title{On-chip parametric amplification with 26.5~dB gain at telecommunication wavelengths  using CMOS-compatible hydrogenated amorphous silicon waveguides}
\author{Bart Kuyken$^{1*}$, St\'ephane Clemmen$^{2*}$, Shankar Kumar Selvaraja$^1$, Wim Bogaerts$^1$, Dries Van Thourhout$^1$, Philippe Emplit$^3$, Serge Massar$^2$, Gunther Roelkens$^1$ and Roel Baets$^{1\dagger}$}
\address{$^1$ Photonics Research Group, INTEC-department, Ghent University - IMEC,  Sint-Pietersnieuwstraat 41, 9000 Gent, Belgium
\\
$^2$ Laboratoire d'information quantique (LIQ), Universit\'e Libre de Bruxelles (U.L.B.), \\ 50~Avenue F. D. Roosevelt, CP 225, B-1050 Bruxelles, Belgium
\\
$^3$ Service OPERA-\textit{photonique}, Universit\'e Libre de Bruxelles (U.L.B.), \\ 50~Avenue F. D. Roosevelt, CP 194/5, B-1050 Bruxelles, Belgium
\\
*Corresponding authors contributed equally to this work~: Bart.Kuyken@Intec.ugent.be \& sclemmen@ulb.ac.be}
%
\begin{abstract}
We present the first study of parametric amplification in hydrogenated amorphous silicon waveguides. Broadband on/off amplification up to 26.5~dB at telecom wavelength is reported. Measured nonlinear parameter is 770~$\textrm{W}^{-1} \textrm{m}^{-1}$, nonlinear absorption 28~$\textrm{W}^{-1} \textrm{m}^{-1}$, bandgap $1.61$~eV. 
\end{abstract}
]
\begin{figure}[t]
\psfrag{xlabel2}[ct][cb][1][0]{\footnotesize{Wavelength (nm)}}
\psfrag{ylabel2}[cb][ct][1][0]{\footnotesize{Peak power (W)}}

\psfrag{1500}[cc][cc][1][0]{\footnotesize{1500}}
\psfrag{1525}[cc][cc][1][0]{\footnotesize{1525}}
\psfrag{1550}[cc][cc][1][0]{\footnotesize{1550}}
\psfrag{1575}[cc][cc][1][0]{\footnotesize{1575}}
\psfrag{1}[cc][cc][1][0]{\footnotesize{1}}
\psfrag{2}[cc][cc][1][0]{\footnotesize{2}}
\psfrag{3}[cc][cc][1][0]{\footnotesize{3}}
\psfrag{4}[cc][cc][1][0]{\footnotesize{4}}
\psfrag{5}[cc][cc][1][0]{\footnotesize{5}}
\psfrag{-78}[cc][cc][1][0]{\footnotesize{-78}}
\psfrag{-76}[cc][cc][1][0]{\footnotesize{-76}}
\psfrag{-74}[cc][cc][1][0]{\footnotesize{-74}}
\psfrag{-72}[cc][cc][1][0]{\footnotesize{-72}}
\psfrag{-70}[cc][cc][1][0]{\footnotesize{-70}}
\psfrag{-68}[cc][cc][1][0]{\footnotesize{-68}}
\psfrag{-66}[cc][cc][1][0]{\footnotesize{-66}}
\psfrag{-64}[cc][cc][1][0]{\footnotesize{-64}}
\psfrag{ylabel1}[cc][cc][1][90]{\footnotesize{dBm/nm}}
%
\centerline{\includegraphics[width=9cm]{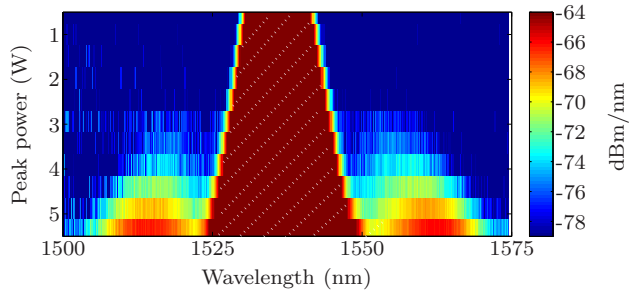}}
\caption{(Color online) Modulation Instability spectra as a function of the  input power which is outside the range of the colour coding at approximately -30~dBm/nm.
\label{fig:MIspectra} 
}
\end{figure}

An important goal in photonics is to realise high performance parametric amplifiers in integrated waveguide circuits. Ideally, this requires low loss waveguides manufactured in materials with high Kerr nonlinearity, operating in the telecommunication window, realized using a CMOS compatible fabrication platform. In this context crystalline silicon (c-Si) waveguides have been studied extensively. However their performance is limited by nonlinear absorption, resulting in wavelength conversion and parametric amplification gains limited to $+5.2$~dB and $+4.2$~dB respectively~\cite{foster-broad-band-2006} in this wavelength range.
In order to overcome this problem, a first possibility is to use a cladding layer with a high figure of merit~\cite{koos-all-optical-2009,bogatscher-2009}  (FOM is the ratio of the nonlinear response to the nonlinear absorption: $\textrm{FOM}=\textrm{Re}[ \gamma ] / (4 \pi \textrm{Im} [ \gamma] )$ with $\gamma$ the nonlinear parameter of the waveguide) but the required evanescent coupling implies stringent design constraints that precludes from good phase matching.
A second approach is for the waveguide core itself to consist of a material with good FOM~\cite{lipson-CMOS-oscillator-2009,razzari-CMOS-OPO-2010,eggleton-net-gain-2008}  but these realizations suffer either from lower linear and nonlinear refractive index or from incompatibility with CMOS fabrication.
One can also step away from the telecommunication wavelength range and work above the c-Si two-photon absorption threshold wavelength of 2.2~$\mu$m~\cite{liu-mid-infrared-2010}. While this approach effectively improves the FOM of the c-Si waveguides, the use for telecommunication applications is not straightforward. 
\\
Silicon-on-insulator (SOI) waveguides with a hydrogenated amorphous silicon (a-Si:H) waveguide core are an alternative to the standard crystalline SOI high-index contrast waveguide platform. 
An appealing feature of this solution is that a-Si:H is deposited at relatively low temperatures and can thus be integrated on a finished CMOS wafer in the back-end.  Recently, the nonlinear response of a-Si:H waveguides was studied~\cite{shoji-ultrafast-2010,preble-optical-2010} and the measured FOM was no better (0.40-0.66)  than that of c-Si waveguides (0.4-0.7~\cite{tsang-nonlinear-2008}). 
Building on our previous work on low-loss a-Si:H waveguides~\cite{selvaraja-low-loss-2009, selvaraja-fabrication-2009}, we present in this paper engineered a-Si:H waveguides with a FOM of $2.1 \pm 0.4$ at telecommunication wavelengths. 
 A similarly high FOM-value was reported in~\cite{Peacock-fiber-2010} in an amorphous silicon core
fiber. The difference with previous  work~\cite{shoji-ultrafast-2010,preble-optical-2010} is presumably due to the fact that a-Si:H is a material with considerable freedom in the chemical structure. Therefore different fabrication procedures can give rise to different material properties. The high linear refractive index of a-Si:H allows high optical confinement, which not only enhances the nonlinear response  of the waveguide but also allows for dispersion engineering. By carefully designing the dimensions of the a-Si:H waveguide it is possible to bring the dispersion of the waveguide into the anomalous regime, which allows for nonlinear effects such as soliton propagation and Modulation Instability (MI)~\cite{agrawal}. In this paper we make use of the latter effect to construct a nonlinear parametric amplifier. 
\\
The circuit is built in 220~nm thick hydrogenated amorphous silicon deposited on top of a 1950~nm thick polished high density silicon dioxide layer.   The 220~nm thick a-Si:H film is deposited by plasma enhanced chemical vapor deposition.  The film was formed using silane (SiH4) as a precursor gas along with helium (He) dilution. In order to achieve low-losses, we used a gas ratio (He/SiH4) of 9, and a plasma power of 180~Watts at a pressure of 2.6~Torr. These typical parameters were obtained from extensive process optimizations~\cite{selvaraja-low-loss-2009}. After forming the waveguide core layer, 500~nm wide photonic wires were patterned using CMOS fabrication technology~\cite{selvaraja-fabrication-2009}. 
The band gap of our a-Si:H films is measured using spectroscopic ellipsometry (in the 300-1600~nm range). The optical constants were extracted from a multilayer model using the Cody-Lorentz model~\cite{cody-disorder} and the recommendations of Ferlauto et al.~\cite{ferlauto-analytical-2002}. In order to have an accurate result, the interface roughness was also taken into account as well as the film thickness measured independently by cross-section scanning electron microscopy. By using seven fitting parameters we obtained a band gap of $1.613 \pm 0.022$~eV (while c-Si has a gap of $1.12$~eV).  The half band gap is thus approximately 0.8~eV, corresponding to $\lambda\simeq1550$~nm.  Working around the half band gap gives rise to a high FOM. 
The single-mode waveguides we use are 500~nm wide and TE polarized light is used in the experiments. In-coupling and out-coupling of light is realized using diffractive grating couplers inducing each 7.5-8~dB loss depending on the wavelength. 
Waveguides with a variety of lengths were fabricated. Except where mentioned explicitly, all results are for waveguides with a length of 1.1~cm.  The fiber to fiber loss was measured for different waveguide lengths ranging from 6~mm to 6~cm at low input powers to extract the linear loss and the incoupling and outcoupling loss, which are assumed to be identical. 
The linear waveguide propagation loss at 1535~nm is 4.8~dB/cm.  The dispersion of the waveguides was estimated from 4-wave-mixing-based conversion efficiency spectrum we measured at low power. Indeed, its bandwidth only depends on the propagation length and the group velocity dispersion $|\beta_2|= 2.0 \textrm{ ps}^2 \textrm{m}^{-1}$. The negative sign of $\beta_2$  (anomalous regime) is deduced from the existence of MI (see below). These values are confirmed by numerical simulation using the commercial software Fimmwave~\copyright \: which predicts $\beta_2=-2.6 \textrm{ ps}^2 \textrm{m}^{-1}$. The imaginary part of the nonlinear parameter, estimated by measuring the nonlinear dependence of the absorption, is $\textrm{Im} [ \gamma ]=28 \pm 3 \textrm{W}^{-1} \textrm{m}^{-1}$. The real part of the nonlinear parameter, determined by comparing the dependence of the XPM fringes on input power with a simulation which takes into account dispersion, nonlinear absorption and presence of free carriers, is $\textrm{Re}[ \gamma ] = 770 \pm 100 \textrm{W}^{-1} \textrm{m}^{-1}$, in agreement with the value obtained by studying photon pair production in these waveguides~\cite{clemmen-photon-2010}. Assuming a nonlinear modal area $A_{\textrm{eff}}=0.07 \: \mu\textrm{m}^2$, the nonlinear index  is $n_2=(1.3\pm0.2 ) \times 10^{-17} \: \textrm{m}^2/ \textrm{W}$. 
During our measurements, we found that the nonlinear properties of our waveguides degrades with time, rapidly during the first minutes and then more slowly.  Annealing the sample at $200^{\circ}$C for $1/2$ hour restores the FOM to its initial value (and this can be done multiple times). This suggests that with proper further optimization of the a-silicon material thermal excitation can compensate for photon-induced decay at moderate
operating temperatures, possibly down to room temperature. All results reported in this paper are taken after a few hours of operation, when the properties of the waveguides change slowly, and without annealing the sample. We intend to report on the decay properties, for different temperatures, in more detail in a future publication. 
\\
Parametric amplification occurs in waveguides with Kerr nonlinearity and anomalous dispersion. When the phase matching conditions are satisfied, a small signal detuned  from the pump frequency by a frequency difference $\Omega$ is exponentially amplified, while simultaneously an idler signal is created at detuning $- \Omega$~\cite{agrawal}. This process can be used both for signal amplification and frequency conversion. 
\begin{figure}[t]
\psfrag{laser}[cc][cc][1][0]{\footnotesize{4~ps}}
\psfrag{edfa}[cc][cc][1][0]{\footnotesize{edfa}}
\psfrag{p-c}[cc][cc][1][0]{\footnotesize{pc}}
\psfrag{w-d}[cc][cc][1][0]{\footnotesize{a-Si:H}}
\psfrag{d-e}[cc][cc][1][0]{\footnotesize{delay}}
\psfrag{t-n}[lc][lc][1][0]{\footnotesize{tunable filter}}
\psfrag{t-f}[cc][cc][1][0]{\footnotesize{filter}}
\psfrag{t-r}[cc][cc][1][0]{\footnotesize{translation}}
\psfrag{Atn}[cc][cc][1][0]{\footnotesize{Attenuator}}
\psfrag{OSA}[cc][cc][1][0]{\footnotesize{OSA}}
\psfrag{lam}[cc][cc][1][0]{\footnotesize{1535~nm}}
\psfrag{mg4}[cc][cc][1][0]{\footnotesize{OSA}}
\psfrag{k-2}[cc][cc][1][0]{Pump}
\psfrag{k-5}[cc][cc][1][0]{Signal}
\psfrag{kk}[cc][cc][1][0]{\footnotesize{a)}}
\psfrag{k-1}[cc][cc][1][0]{\footnotesize{a)}}
\psfrag{v-0}[cb][cb][1][0]{\footnotesize{1500}}
\psfrag{v-1}[cb][cb][1][0]{\footnotesize{1540}}
\psfrag{v-2}[cb][cb][1][0]{\footnotesize{1580}}
\psfrag{t-0}[cb][cb][1][0]{\footnotesize{-80}}
\psfrag{t-1}[cb][cb][1][0]{\footnotesize{-60}}
\psfrag{t-2}[cb][cb][1][0]{\footnotesize{-40}}
\psfrag{xlabel1}[cc][cb][1][0]{\footnotesize{Wavelength (nm)}}
\psfrag{Z}[cc][cb][1][90]{\footnotesize{Spectrum (dB)}}

\psfrag{k-3}[cc][cb][1][0]{\footnotesize{b)}}
\psfrag{k-4}[cc][cb][1][0]{\footnotesize{\color{red}Signal}}
\centerline{\includegraphics[width=8.5cm]{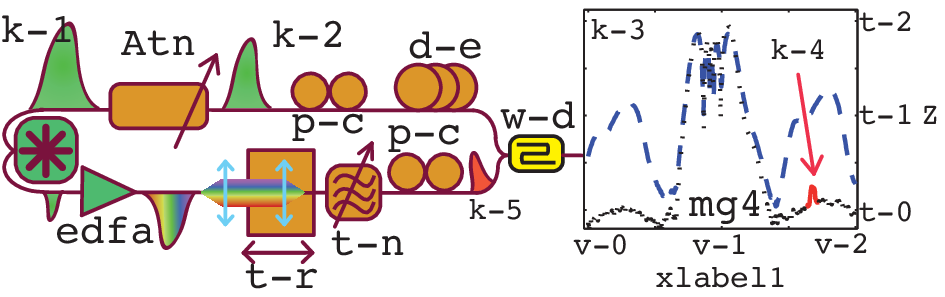}}
\psfrag{xlabel2}[ct][cc][1][0]{\footnotesize{Power (W)}}
\psfrag{ylabel2}[cb][cc][1][0]{\footnotesize{Amplification (dB)}}
\psfrag{xlabel4}[ct][cc][1][0]{\footnotesize{Wavelength (nm)}}
\psfrag{ylabel4}[ct][cc][1][0]{\footnotesize{Amplification (dB)}}
\psfrag{v0}[cb][cb][1][0]{\footnotesize{0}}
\psfrag{v1}[cb][cb][1][0]{\footnotesize{1}}
\psfrag{v2}[cb][cb][1][0]{\footnotesize{2}}
\psfrag{v3}[cb][cb][1][0]{\footnotesize{3}}
\psfrag{v4}[cb][cb][1][0]{\footnotesize{4}}
\psfrag{v5}[cb][cb][1][0]{\footnotesize{5}}
\psfrag{v6}[cb][cb][1][0]{\footnotesize{6}}
\psfrag{w0}[cc][ct][1][0]{\footnotesize{0}}
\psfrag{w1}[cc][cc][1][0]{\footnotesize{5}}
\psfrag{w2}[cc][cc][1][0]{\footnotesize{10}}
\psfrag{w3}[cc][cc][1][0]{\footnotesize{15}}
\psfrag{w4}[cc][cc][1][0]{\footnotesize{20}}
\psfrag{w5}[cc][cc][1][0]{\footnotesize{25}}
\psfrag{w6}[cc][cc][1][0]{\footnotesize{30}}

\psfrag{x0}[cb][cb][1][0]{\footnotesize{1540}}
\psfrag{x1}[cb][cb][1][0]{\footnotesize{1560}}
\psfrag{x2}[cb][cb][1][0]{\footnotesize{1580}}
\psfrag{x3}[cb][cb][1][0]{\footnotesize{1560}}
%
\psfrag{z0}[cb][cb][1][0]{\footnotesize{1540}}
\psfrag{z1}[cb][cb][1][0]{\footnotesize{1560}}
\psfrag{z2}[cb][cb][1][0]{\footnotesize{1580}}
\psfrag{z3}[cb][cb][1][0]{\footnotesize{1600}}
\psfrag{y0}[cc][ct][1][0]{\footnotesize{0}}
\psfrag{y1}[cc][cc][1][0]{\footnotesize{5}}
\psfrag{y2}[cc][cc][1][0]{\footnotesize{10}}
\psfrag{y3}[cc][cc][1][0]{\footnotesize{15}}
\psfrag{y4}[cc][cc][1][0]{\footnotesize{20}}
\psfrag{y5}[cc][cc][1][0]{\footnotesize{25}}
\psfrag{y6}[cc][cc][1][0]{\footnotesize{30}}
%
\psfrag{bb}[cc][cc][1][0]{\footnotesize{d)}}
\psfrag{cc}[cc][cc][1][0]{\footnotesize{c)}}
\psfrag{mm}[cc][cc][1][0]{\footnotesize{\color{red}Signal}}
\centerline{\includegraphics[width=8cm]{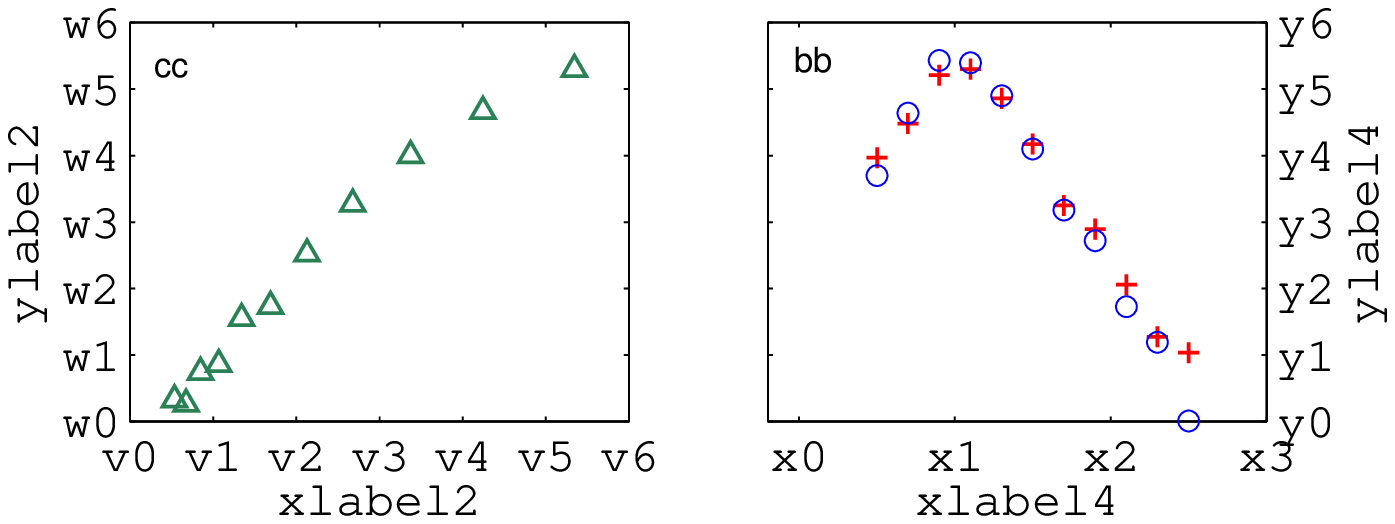}}
\caption{(Color online) Pump-probe experiment.  
a) The experimental setup combinesan intense pump pulse with a signal pulse whose frequency and time delay can be adjusted. 
b) When the pump {\scriptsize{($\cdots$)}} and signal {\scriptsize{({\color{red}-})}} pulses are not synchronized, the signal is very small in comparison to the pump pulse. If both pulses are synchronized {\scriptsize{({\color{blue}$--$})}}, signal pulses are amplified by more than 20~dB and are converted into idler pulses. Note that the pump pulse is broadened spectrally due to self phase modulation, which in turn induces a broadening of the signal pulse.
c) On/off gain as a function of peak power {\scriptsize{({\color{green}$\Delta$})}}.
d) On/off gain (resp. frequency conversion efficiency) as a function of wavelength {\scriptsize{({\color{red}+}}}, resp. {\scriptsize{\color{blue}$\circ$})}. 
\label{fig:amplification}
}
\end{figure} 
We studied parametric amplification in a-Si:H waveguide using a pump-probe experiment, see Fig.~\ref{fig:amplification}. The pump was produced by a fiber laser emitting pulses at 1535~nm with a spectral width of 0.67~nm (FWHM), a duration of 3.8~ps (FWHM) at 10~MHz repetition rate. The peak power in the waveguide was adjusted from 0 to 5.3~W  thanks to a variable attenuator. Signal pulses are obtained from the secondary output of the laser: they are spectrally broadened using the optical non linearities in an EDFA, and then selected by a 1.2~nm tunable passband filter to generate  (signal) pulses at arbitrary wavelengths ranging from 1550~nm to 1590~nm. The coupled peak power of the signal was always kept below 100~$\mu$W.  An optical delay line was used to synchronize or desynchronize the signal and the pulse. After passing polarization controllers the combined pump and signal were coupled into the waveguide.  At the output of the waveguide, the light is  sent to an optical spectrum analyzer.
%
The parametric gain as a function of input pump power and signal wavelength was calculated after removing the effect of spontaneous MI. This was done by subtracting the MI spectrum when signal and pump are not synchronized from the output spectrum when they are synchronized. The on/off gain was calculated in a similar way as in~\cite{liu-mid-infrared-2010} by integrating the power in the sideband caused by the amplification of the signal pulse. We obtained a maximum on/off gain of 26.5~dB and an on/off conversion efficiency of 27~dB for a signal at 1562~nm and an on-chip pump peak power of 5.3~W.  Taking into account propagation losses, this corresponds to a net on-chip amplification of 21.2~dB, resulting in enough gain to overcome the high incoupling and outcoupling losses, giving rise to 6.2~dB net off-chip amplification. The amplification as a function of the pump power at wavelength $1562$~nm is plotted in Fig.~\ref{fig:amplification}.c.  This amplification is to be compared with the $4.2$~dB on/off amplification that was observed in crystalline silicon waveguide structures~\cite{foster-broad-band-2006}. 
Fig.~\ref{fig:amplification}.d  shows  gain and conversion efficiency spectra for the peak pump power of 5.3~W. On/off gain in excess of 20~dB was observed in the band 1550-1570~nm, and on/off gain in excess of 10~dB was observed in the band 1550-1582~nm. 
\\
In the absence of a signal beam, the quasi-continuous nature of the pump pulse is broken by spontaneous MI~\cite{agrawal}. The spectra when only the pump pulse travels through the a-Si:H waveguide are shown in Fig.~\ref{fig:MIspectra} for various pump power levels. The bandwidth and strength of the MI sidebands  increase with coupled peak input power, while the pump itself also broadens due to self phase modulation. Note that MI peak can be related to amplification by comparison with the spectral density in absence of MI. Spectral density at the MI-peak wavelength before a-Si:H chip is less than $-80$~dB/nm and experiences $20$~dB fiber-to-fiber loss afterwards. Typical value of MI peak, above $-70$~dBm/nm, should thus be compared with background level below $-100$~dB/nm, giving an amplification significantly greater than $30$~dB. The difference with the maximum gain reported in Fig.~\ref{fig:amplification}.c is presumably that in the pump probe experiment, at maximum power, the pump starts to become depleted which limits the gain.
%

In summary, CMOS compatible hydrogenated amorphous silicon waveguides provide on chip parametric amplification at telecommunication wavelengths. Optimisation of the fabrication process could lead to further increases in the nonlinear figure of merit. 
%

B.K. thanks W. Green and X. Liu for helpfull discussions. S.C. acknowledges the support of the Fonds pour la formation {\`a} la Recherche dans l'Industrie et dans l'Agriculture (FRIA, Belgium). B.K., W.B. and G.R. acknowledge the Flemish Research Foundation (FWO-Vlaanderen) for a (post) doctoral fellowship. All authors acknowledge support by  the Interuniversity Attraction Poles Photonics@be Programme (Belgian Science Policy) under grant IAP6-10. 
\end{document}